%
\newcount\mgnf\newcount\tipi\newcount\tipoformule
\newcount\aux\newcount\casa\newcount\piepagina
\mgnf=0          
\aux=0           
           \def\9#1{\ifnum\aux=1#1\else\relax\fi}
\tipoformule=1   
\piepagina=1     
\ifnum\piepagina=0
\footline={\rlap{\hbox{\copy200}\ $\st[\number\pageno]$}\hss\tenrm
\foglio\hss}\fi
\ifnum\piepagina=1
\footline={\rlap{\hbox{\copy200}} \hss\tenrm
\folio\hss}\fi
\ifnum\piepagina=2\footline{\hss\tenrm\folio\hss}\fi
\newcount\referenze
\referenze=0         
\ifnum\referenze=0
\openout7=\jobname.b\def\[#1]{[#1]\write7{[#1]}}\fi
\ifnum\referenze=1
\input \jobname.bb \fi

\casa=0          
\ifnum\mgnf=0
   \magnification=\magstep0
\hsize=12truecm\vsize=19.5truecm
   \parindent=4.pt\fi
\ifnum\mgnf=1
   \magnification=\magstep1
\hsize=16.0truecm\vsize=22.5truecm\baselineskip14pt\vglue6.3truecm
\overfullrule=0pt \parindent=4.pt\fi
%
\let\a=\alpha       \let\d=\delta 
     \let\th=\vartheta \let\l=\lambda
\let\m=\mu                   
\let\s=\sigma     
   
     \let\L=\Lambda
            \let\O=\Omega

{\count255=\time\divide\count255 by 60 \xdef\oramin{\number\count255}
        \multiply\count255 by-60\advance\count255 by\time
   \xdef\oramin{\oramin:\ifnum\count255<10 0\fi\the\count255}}
\def\ora{\oramin }

\def\data{\number\day/\ifcase\month\or gennaio \or febbraio \or marzo \or
aprile \or maggio \or giugno \or luglio \or agosto \or settembre
\or ottobre \or novembre \or dicembre \fi/\number\year;\ \ora}

\setbox200\hbox{$\scriptscriptstyle 27/febbraio/96 $}

\newcount\pgn \pgn=1
\def\foglio{\number\numsec:\number\pgn
\global\advance\pgn by 1}
\def\foglioa{A\number\numsec:\number\pgn
\global\advance\pgn by 1}


\global\newcount\numsec\global\newcount\numfor
\global\newcount\numfig
\gdef\profonditastruttura{\dp\strutbox}
\def\senondefinito#1{\expandafter\ifx\csname#1\endcsname\relax}
\def\SIA #1,#2,#3 {\senondefinito{#1#2}
\expandafter\xdef\csname #1#2\endcsname{#3} \else
\write16{???? ma #1,#2 e' gia' stato definito !!!!} \fi}
\def\etichetta(#1){(\veroparagrafo.\veraformula)
\SIA e,#1,(\veroparagrafo.\veraformula)
 \global\advance\numfor by 1
\9{\write15{\string\FU (#1){\equ(#1)}}}
\9{ \write16{ EQ \equ(#1) == #1  }}}
\def \FU(#1)#2{\SIA fu,#1,#2 }
\def\etichettaa(#1){(A\veroparagrafo.\veraformula)
 \SIA e,#1,(A\veroparagrafo.\veraformula)
 \global\advance\numfor by 1
\9{\write15{\string\FU (#1){\equ(#1)}}}
\9{ \write16{ EQ \equ(#1) == #1  }}}
\def\getichetta(#1){Fig. \verafigura
 \SIA e,#1,{\verafigura}
 \global\advance\numfig by 1
\9{\write15{\string\FU (#1){\equ(#1)}}}
\9{ \write16{ Fig. \equ(#1) ha simbolo  #1  }}}
\newdimen\gwidth
\def\BOZZA{
\def\alato(##1){
 {\vtop to \profonditastruttura{\baselineskip
 \profonditastruttura\vss
 \rlap{\kern-\hsize\kern-1.2truecm{$\scriptstyle##1$}}}}}
\def\galato(##1){ \gwidth=\hsize \divide\gwidth by 2
 {\vtop to \profonditastruttura{\baselineskip
 \profonditastruttura\vss
 \rlap{\kern-\gwidth\kern-1.2truecm{$\scriptstyle##1$}}}}}
} \def\alato(#1){} \def\galato(#1){}
\def\veroparagrafo{\number\numsec}\def\veraformula{\number\numfor}
\def\verafigura{\number\numfig} \def\geq(#1){\getichetta(#1)\galato(#1)}
\def\Eq(#1){\eqno{\etichetta(#1)\alato(#1)}}
\def\eq(#1){\etichetta(#1)\alato(#1)}
\def\Eqa(#1){\eqno{\etichettaa(#1)\alato(#1)}}
\def\eqa(#1){\etichettaa(#1)\alato(#1)}
\def\eqv(#1){\senondefinito{fu#1}$\clubsuit$#1\write16{No translation
for #1}%
\else\csname fu#1\endcsname\fi}
\def\equ(#1){\senondefinito{e#1}\eqv(#1)\else\csname e#1\endcsname\fi}

\openin13=#1.aux \ifeof13 \relax \else
\input #1.aux \closein13\fi
\openin14=\jobname.aux \ifeof14 \relax \else
\input \jobname.aux \closein14 \fi
\9{\openout15=\jobname.aux}
\newskip\ttglue
\ifnum\casa=1
\font\dodiciit=cmr12
\font\titolo=cmr12
\else
\font\dodiciit=cmti12
\font\titolo=cmbx12 scaled \magstep2
\fi
\font\ottorm=cmr8\font\ottoi=cmmi8\font\ottosy=cmsy8
\font\ottobf=cmbx8\font\ottott=cmtt8\font\ottosl=cmsl8\font\ottoit=cmti8
\font\sixrm=cmr6\font\sixbf=cmbx6\font\sixi=cmmi6\font\sixsy=cmsy6
\font\fiverm=cmr5\font\fivesy=cmsy5\font\fivei=cmmi5\font\fivebf=cmbx5
\def\ottopunti{\def\rm{\fam0\ottorm}
\textfont0=\ottorm \scriptfont0=\sixrm \scriptscriptfont0=\fiverm
\textfont1=\ottoi \scriptfont1=\sixi   \scriptscriptfont1=\fivei
\textfont2=\ottosy \scriptfont2=\sixsy   \scriptscriptfont2=\fivesy
\textfont3=\tenex \scriptfont3=\tenex   \scriptscriptfont3=\tenex
\textfont\itfam=\ottoit  \def\it{\fam\itfam\ottoit}%
\textfont\slfam=\ottosl  \def\sl{\fam\slfam\ottosl}%
\textfont\ttfam=\ottott  \def\tt{\fam\ttfam\ottott}%
\textfont\bffam=\ottobf  \scriptfont\bffam=\sixbf
\scriptscriptfont\bffam=\fivebf  \def\bf{\fam\bffam\ottobf}%
\tt \ttglue=.5em plus.25em minus.15em
\setbox\strutbox=\hbox{\vrule height7pt depth2pt width0pt}%
\normalbaselineskip=9pt\let\sc=\sixrm \normalbaselines\rm}
\catcode`@=11      
\def\footnote#1{\edef\@sf{\spacefactor\the\spacefactor}#1\@sf
\insert\footins\bgroup\ottopunti
 \interlinepenalty100 \let\par=\endgraf
   \leftskip=0pt \rightskip=0pt
   \splittopskip=10pt plus 1pt minus 1pt \floatingpenalty=20000
   \smallskip\item{#1}\bgroup\strut\aftergroup\@foot\let\next}
\skip\footins=12pt plus 2pt minus 4pt
\dimen\footins=30pc
\catcode`@=12
\newdimen\xshift \newdimen\xwidth \newdimen\yshift

\def\ins#1#2#3{\vbox to0pt{\kern-#2 \hbox{\kern#1 #3}\vss}\nointerlineskip}

\def\eqfig#1#2#3#4#5{
\par\xwidth=#1 \xshift=\hsize \advance\xshift
by-\xwidth \divide\xshift by 2
\yshift=#2 \divide\yshift by 2
\line{\hglue\xshift \vbox to #2{\vfil
#3 \includegraphics{#4.ps}
}\hfill\raise\yshift\hbox{#5}}}


\def\8{\write13}


\def\V#1{{\,\underline#1\,}}
\def\T#1{#1\kern-4pt\lower9pt\hbox{$\widetilde{}$}\kern4pt{}}

\let\io=\infty\let\ig=\int
\def\fra#1#2{{#1\over#2}}\let\0=\noindent

\def\guida{\leaders\hbox to 1em{\hss.\hss}\hfill}
\def\tende#1{\vtop{\ialign{##\crcr\rightarrowfill\crcr
              \noalign{\kern-1pt\nointerlineskip}
              \hglue3.pt${\scriptstyle #1}$\hglue3.pt\crcr}}}
\def\otto{\ {\kern-1.truept\leftarrow\kern-5.truept\to\kern-1.truept}\ }

\def\pagina{\vfill\eject}
\let\ciao=\bye

\def\st{\scriptscriptstyle}
\def\*{\vskip0.3truecm}
\def\lis#1{{\overline #1}}
\def\eg{\hbox{\it e.g.\ }}
\def\ap{\hbox{\it a priori\ }}
\def\ie{\hbox{\it i.e.\ }}
\def\fiat{{}}

\def\CC{{\cal C}}

\def\\{\hfill\break}
\def\={{ \; \equiv \; }}

\fiat

\def\PP{{\cal P}}

\def\annota#1#2{\footnote{${}^#1$}{{\parindent=0pt\rm
\0#2\vfill}}}

\ifnum\aux=1\BOZZA\else\relax\fi
\ifnum\tipoformule=1\let\Eq=\eqno\def\eq{}\let\Eqa=\eqno\def\eqa{}
\def\equ{{}}\fi


\def\pallino{{\0$\bullet$}}\def\1{\ifnum\mgnf=0\pagina\else\relax\fi}
\fiat

\vskip.5cm
\centerline{\titolo Reversibility, coarse graining and}
\vskip.5cm
\centerline{\titolo the chaoticity principle\annota{*}{{\it
mp$\_$arc@math.utexas.edu}, \#96-50, and
{\it chao-dyn@xyz.lanl.gov},\#9512??.}}
\vskip1.cm
\centerline{\dodiciit F. Bonetto, G. Gallavotti}
\centerline{Fisica, Universit\`a
di Roma La Sapienza, P.le Moro 2, 00185, Roma, Italia.}

\vskip1cm

\0{\it Abstract: We describe a way of interpreting  the chaotic
principle of {\rm [GC1]} more extensively than it was meant in the
original works. Mathematically the analysis is based on the dynamical
notions of Axiom A and Axiom B and on the notion of Axiom C, that we
introduce arguing that it is suggested by the results of an experiment
([BGG]) on chaotic motions. Physically we interpret a breakdown of the
Anosov property of a time reversible attractor (replaced, as a control
parameter changes, by an Axiom A property) as a spontaneous breakdown
of the time reversal symmetry: the relation between time reversal and
the symmetry that remains after the breakdown is analogous to the
breakdown of $T$-invariance while $TCP$ still holds.} \*

\0{\it Keywords}: {\sl Strange attractors, Chaoticity principle, Time
reversal, TCP}

\vskip.5cm
\*

\0{\bf\S1: Introduction.}
\numsec=1\numfor=1\pgn=1
\*

In reference \[GC2] a general mechanical system in a non equilibrium
situation was considered. Calling $\CC$ the (compact or "finite") phase
space of the "observed events", $\m_0$ the volume measure on it and $S$
the map describing the time evolution (regarded as a discrete invertible
mapping of "observed events" into the "next ones", \ie as a Poincar\'e's
map on some surface in the full phase space) a principle holding when
motions have an empirically chaotic nature was introduced:

\*
{\it Chaotic hypothesis: A chaotic many particle system in a stationary
state can be regarded, for the purpose of computing macroscopic
properties, as a smooth dynamical system with a transitive Axiom A
globally attracting set.  In reversible systems it can be regarded, for
the same purposes, as a smooth transitive Anosov system.} \*

\pallino{\it Chaotic} is an empirical qualitative notion that means
that most points of the attracting set have a stable and an unstable
manifold with positive dimension.  In the applications in [GC1], [GC2],
[G1] the use of the hypothesis, which is a natural extension of a
principle proposed by Ruelle, was based on reversibility and on
transitivity.

\pallino An {\it attracting set} is a closed invariant set such that
all points in its vicinity evolve (in the future) tending to it and
such that no subset has the same property (\ie it is "minimal").  A set
is {\it globally attracting} if it is attracting and {\it all} points
of an open dense set evolve tending to it.

\pallino{\it Axiom A} attracting set is a hyperbolic
attracting set: \ie an attracting set with each of
its points possessing stable and unstable manifolds depending
coninuously upon the points and with contraction and expansion rates
bounded uniformly away from $0$. Furthermore the periodic points are dense.

\pallino{\it Reversibility} means that there is a map $i$ of $\CC$ onto
itself that changes the sign to time in the sense that $iS=S^{-1}i$ and
$i^2=1$. As is well known reversibility should not be confused with the
invertibility of the map $S$ (always assumed below).

\pallino{\it Transitivity} is intended to mean that the stable and
unstable manifolds of the attracting set points are dense on it (this
is not a very strong requirement in view of (7.6) in [Sm], p.  783).

\pallino{\it Anosov system} is a dynamical system in which the whole
phase space is hyperbolic.
\*

Note that if the Axiom A attracting set is supposed to be also a smooth
manifold then the restriction of the dynamics to it an Anosov
system and this is the meaning that we give, in this paper, to the
second assumption in the above hypothesis (see \S6 of [BGG]).

However in the previous paper [G4] transitivity was instead intended to
mean, at least in the reversible cases, density of the stable and
unstable manifolds of the attracting set points on the entire phase space
(so that the system was in fact a transitive Anosov system).  This is
not always a property that one may be willing to consider as
reasonable.

It is reasonable for systems that are very close to conservative ones
(as in [G4]).  But it is very likely (see [BGG], \S6, Fig. 14) to be
incorrect in systems that are under strong non conservative forces,
even if still evolving with a reversible dynamics.  In fact the
attracting sets of such systems often evolve, as the strength of the
forces increases, from a very chaotic initial attracting set to a more
ordered situation characterized by a periodic orbit or by a very small
attracting "tube" almost identical to a periodic orbit; the evolution
shows a gradual decrease of the dimension and of the phase space region
occupied by the attracting set, which therefore quite soon may become
contained in a proper closed subset of phase space (so that the system
cannot have stable and unstable manifolds with dimension half that of
phase space: a necessary consequence of time reversibility in
transitive Anosov systems).

In such cases it is still reasonable, see [R1], [ER], to think that the
attracting set, if chaotic, can be regarded "just" as an {\it
Axiom A attracting set}, an assumption weaker than assuming that the system
is an Anosov system.

What can be said for such systems? is there a suitable reformulation of
the chaoticity principle that could make it applicable even in very
strongly forced (but still "chaotic") systems making possible an
analysis similar to that leading to the fluctuation theorem of [GC1],
[GC2] or to the Onsager relations of [G4]? These are the questions we
address here.  \*

\0{\it\S2. A distinction between Anosov and Axiom A properties.
Statistics. Axioms B and C.}
\numsec=2\numfor=1\*

We shall try to adopt the notations used in the well known  paper by
Smale, [Sm].

Furthermore in this paper, as in [GC1],[GC2], a distinction will be made
between an {\it attractor} and its closure that we call more properly
an {\it attracting set}: this is often a rather
confusing point because some authors identify an attractor with its
closure. Here we shall not.
Thus, recalling that $\m_0$ denotes the volume measure on $\CC$, we
shall formally say that:
\*

\0{\bf Definition 1:} {\it A point $x\in\CC$ admits a statistics if there
is a probability distribution in phase space such that:
$$\lim_{M\to\io}\fra1M\sum_{j=0}^{M-1} F(S^jx)=\ig_\CC
F(y)\m(dy)\Eq(2.1)$$
for all continuous functions $F$ on phase space $\CC$.}
\*

\0and:
\*

\0{\bf Definition 2:} {\it An attractor is a set $C$ enjoying the
properties: (a) it is invariant and dense in an attracting set, (b) it
has minimal Hausdorff dimension and (c) $\m_0$--almost all points in its
vicinity admit the {\sl same statistics $\m$ and $\m(C)=1$}, (d) it is 
minimal.\annota{1}{This means that any other set with the same
properties has a closure that contains the closure ${\rm clos}(C)$.} An
attracting set verifies Axiom A if it has the properties: (i) each of
its points is hyperbolic, (ii) the periodic points are dense on it,
(iii) it contains a dense orbit.}
\*

With this definition one has to live with the fact that there will be
several essentially identical attractors: if the distribution $\m$
gives, for instance, $0$ probability to individual points on the
attractor $C$ then any subset of $C$ obtained by removing a countable
number of points will still be an attractor (of course with the same
closure as $C$ itself).  This is the main reason why it is wise
to distinguish the notions of attractor and of attracting set;
in the cases met in this paper it will be appropriate to call the
latter an {\it attracting basic set} (see below).

It is useful to recall some more general definitions and properties:
\*
\pallino a dynamical
system {\it verifies Axiom A} if each point in the set $\O$ of
"nonwandering points" (\ie in the set of all "recurrent points") is
"hyperbolic", \ie each nonwandering point admits stable and unstable
manifolds, continuously dependent on the point and with expansion and
contraction rates uniformly bounded away from $0$; furthermore the
periodic points are dense on $\O$, [Sm], p.777.  The closure of an
attractor for an Axiom A system is an Axiom A attracting set in the
above sense.

\pallino Axiom A systems have a rather simple structure as the sets of
their nonwandering points consist of a finite number of closed
invariant indecomposable topologically transitive sets, called {\it
basic sets}, see [Sm], p.777.\annota{2}{Topologically transitive, [Sm]
p.776, means that they contain a point with a dense orbit.  Note that
this is weaker than transitive in the sense of \S 1.  Indecomposable
means that they contain no subset with the same properties.  The basic
sets are the {\it building} pieces (or the "bases") of the part of
phase space where the dynamics is non trivial.  Transitive Anosov
systems are simply Axiom A systems with a (unique) basic set coinciding
with the whole phase space.}

\pallino A basic set that is the closure of an attractor is called an
attracting basic set, likewise one defines a repelling basic
set.\annota{3}{Thus a basic set for an Axiom A system can be regarded
as a dynamical system in itself: in this case it may fail to be Anosov
only because in general it is not a smooth manifold but just a closed
set.}

\pallino Another interesting class of dynamical systems is the class of
{\it Axiom B systems} (see [Sm.] p 778): they are the dynamical systems
verifying Axiom A and the further mild transversality
property\annota{4}{Mild because of the theorem (6.7), p.  779, in
[Sm].} that if $\O_i,\O_j$ are a pair of basic sets such that the
stable set $W^s(\O_i)$ and the unstable set $W^u(\O_j)$ intersect {\it
then} they intersect transversally (see below).

\pallino The stable (resp.  unstable) set of $\O_i$ (resp.  $\O_j$) is
the union of the stable (resp.  unstable) manifolds of all its points
(see [Sm], p.  777).  The intersection between $W^s(\O_i)$ and
$W^u(\O_j)$ is {\it transversal} if the stable manifold $W^s(p)$ and
the unstable manifold $W^u(q)$ of any two {\it periodic points} $p\in
\O_i$ and $q\in\O_j$ intersect transversally.  Finally two manifolds
intersect transversally at a point if their tangent planes span the
full tangent plane (see [Sm], p.  752).
\*

We also need to recall that, particularly in the numerical experiments,
for dynamical systems with "chaotic behavior" it is usually also
assumed that, after the "obvious" conservation laws and symmetries are
taken into account, the {\it whole phase space} admits a statistics, in
the sense that almost all points (with respect to the volume measure)
admit a statistics, the same for all of them. This is usually called the
{\it zero-th law}, [UF]:
\*

\0{\bf Extended zero-th law:} {\it A dynamical system $(\CC,S)$
describing a many particle system (or a continuum such as a fluid)
generates motions that admit a statistics $\m$ in the sense that, given
any (smooth) macroscopic observable $F$ defined on the points $x$ of the
phase space $\CC$, the time average of $F$ exists for all
$\m_0$--randomly--chosen initial data $x$ and is given by:
$$\lim_{M\to\io}\fra1M\sum_{k=0}^{M-1}
F(S^jx)=\ig_\CC\m(dx')F(x')\Eq(2.2)$$
with $\m$ being a $S$--invariant probability distribution on $\CC$.}
\*

It is important to note the physical meaning of the above law: in fact,
among other, it implies that the dynamical system $(\CC,S)$ {\it cannot
have more than one attracting basic set}.  We shall say that a system
for which the property described by the above "law" holds {\it verifies
the $0$--th law}.

For our purposes only systems verifying at least Axiom B will be
relevant. For such systems the notion of {\it diagram} of a dynamical
system, see [Sm] p. 754, allows us to interpret the $0$--th law as
saying that the diagram of the system is a partially ordered
set with unique top and bottom points.

In this paper we shall deal with reversible dynamical systems $(\CC,
S)$ verifying Axiom B and the $0$-th law: hence with a unique
attracting basic set $\Omega_+$ and a unique repelling basic set
$\Omega_-$.  Furthermore the attracting set will be assumed transitive
in the sense that the stable and unstable manifolds of each of its
points are dense on it.

The Axiom B and the zeroth law could be reasonably taken as
definitions of models for globally "chaotic" or "globally hyperbolic
systems".

However the problem that we pose in the next section suggests that the
appropriate notion for "globally hyperbolic" or "globally chaotic"
dynamical systems is somewhat stronger. Its definition has been
suggested by our effort to interpret the results of the experiment [BGG]
and we allowed ourselves to give the name of {\it Axiom C systems} to
systems verifying a stronger property; to describe it we introduce
the notion of distance of a point $x$ to the basic sets $\{\O_i\}$ of
an Axiom A system as:
$$\d(x)=\min\{\min_i \fra{d_{\O_i}(x)}{d_0}, \min_{j,\,-\io<n<+\io}
\fra{d_{\O_j}(S^nx)}{d_0}\}\Eq(2.3)$$
where $d_0$ is the diameter of the phase space $\CC$, $d_{\O_i}(x)$ is
the distance of the point $x$ from the basic set $\Omega_i$, the
minimum over $i$ runs over the attracting or repelling basic sets and
the minimum over $j$ runs over the other $k\ge0$ basic sets.

We can then define the {\it Axiom C} systems as:
\*

\0{\bf Definition 3:} {\it A smooth dynamical system $(\CC,S)$ verifies
Axiom C if it is Anosov or if it verifies Axiom A and:

\0(1) among the basic sets there are a unique attracting and a unique
repelling basic sets, denoted $\O_+,\O_-$ respectively, with (open)
dense basins that we call the {\it poles} of the system (future or
attracting and past or repelling poles, respectively).

\0(2) for every $x\in\CC$ the tangent space $T_x$ admits a
H\"older--continuous\annota{5}{One might prefer to require real
smoothness, \eg $C^p$ with $1\le p\le \io$: but this would be too much
for rather trivial reasons. On the other hand H\"older continuity might
be equivalent to simple $C^0$--continuity as in the case of Anosov
systems, see [AA], [Sm].} decomposition as a direct sum of three
subspaces $T^u_x,T^s_x,T^m_x$ such that:

\halign{\qquad
#\quad&#           &#     \hfill              &\qquad#\hfill\cr
a)& $dS\, T^\alpha_x$&$=T^\alpha_{Sx}$          &$\a=u,s,m$ \cr
b)& $|dS^n w|$     &$\le C e^{-\l n}|w|$,     &$w\in T^s_x,\ n\ge0$\cr
c)& $|dS^{-n} w|$  &$\le C e^{-\l n}|w|$,     &$w\in T^u_x,\ n\ge0$\cr
d)& $|dS^n w|$     &$\le C \d(x)^{-1} e^{-\l |n|}|w|$,&$w\in T^m_x,\
\forall n$\cr}

\0where the dimensions of $T^u_x,T^s_x, T^m_x$ are $>0$ and $\d(x)$ is
defined in \equ(2.3).

\0(3) if $x$ is on the attracting basic set $\O_+$ then $T^s_x\oplus
T^m_x$ is tangent to the stable manifold in $x$; viceversa if $x$ is on
the repelling basic set $\O_-$ then $T^u_x\oplus T^m_x$ is tangent to the
unstable manifold in $x$.} \*

Although $T^u_x$ and $T^s_x$ are not uniquely determined the planes
$T^s_x\oplus T^m_x$ and $T^u_x\oplus T^m_x$ are uniquely determined for
$x\in\O_+$ and, respectively, $x\in\O_-$.

It is clear that an Axiom C system is necessarily also an Axiom B
system verifying the $0$--th law (as it follows from [R2]).  We do not
know an example of an Axiom B system with a unique attracting and a
unique repelling basic set which is not at the same time an Axiom C
system.

Apart from property (1) that is meant to imply the validity of the
$0$--th law,  one can also say that ("at most") the real difference
between an Axiom B and an Axiom C system is that the latter has a
stronger, and more global, hyperbolicity property.

Namely, if $\O_+$ and $\O_-$ are the two poles of the system the stable
manifold of a periodic point $p\in \O_+$ and the unstable manifold of a
periodic point $q\in\O_-$ not only have a point of transversal
intersection, but they intersect transversally {\it all the way} on a
manifold connecting $\O_+$ to $\O_-$; the unstable manifold of a point
in $\O_-$ will accumulate on $\O_+$ {\it without winding around it}.

In fact one can "attach" to $W^s(p)$, $p\in\O_+$, points on $\O_-$ as
follows: we say that a point $z\in \O_-$ is {\it attached} to $W^s(p)$
if it is an accumulation point for $W^s(p)$ and there is a curve
{\it with finite length} linking a point $z_0\in W^s(p)$ to $z$ and {\it
entirely lying on $W^s(p)$}, with the exeption of the endpoint $z$.  A
drawing helps understanding this simple geometrical construction,
slightly unusual because of the density of $W^s(p)$ on $\O_-$,.

We call $\lis W^s(p)$ the set of the points {\it either on $W^s(p)$ or
just attached to $W^s(p)$} on the system basic sets (the set $\lis
W^s(p)$ should not be confused with the closure ${\rm clos}(W^s(p))$,
which is the whole space, see [Sm], p.  783).  If a system
verifies Axiom C the set $\lis W^s(p)$ intersects $\O_-$ on a stable
manifold, by 2) in the above definition.

The definition of $\lis W^u(q)$, $q\in \O_-$, is defined symmetrically by
exchanging $\O_+$ with $\O_-$.

Furthermore if a system verifies Axiom C and $p\in \Omega_+$, $q\in
\Omega_-$ are two periodic points, on the attracting basic set and on
the repelling basic set of the system respectively, then $\lis W^s(p)$
and $\lis W^u(q)$ have a dense set of points in $\Omega_+$ and
$\Omega_-$, respectively.  Furthermore if $z\in C_+$ is one such point
then the surface $\lis W^s(p)\cap\lis W^u(q)$ intersects $C_-$ in a
unique point $\tilde z\=\tilde\imath z$ which can be reached by the
shortest smooth path on $\lis W^s(p)\cap\lis W^u(q)$ linking $z$ to
$C_-$ (the path is on the surface obtained as the envelope of the
tangent planes $T^m$, but it is in general not unique even if $T_m$ has
dimension $1$, see the example in \S4 below).

The map $\tilde\imath$ commutes with both $S$ and $i$, squares to the
identity and will play a key role in the following analysis.

We conjecture that the Axiom C systems are $\O$--stable in the sense of
Smale, [Sm] p.  749.\*

\0{\it\S3. Axiom A, B, C and time reversibility: the problem.}
\numsec=3\numfor=1
\*

If one considers the closure $\O_+={\rm clos}(C)$ of an attractor $C$
verifying Axiom A then the action of the dynamics
$S$ on it fails to be an Anosov system only because ${\rm clos}(C)$
{\it might} be a fractal set rather than a smooth surface.

In nonequilibrium statistical mechanics the dimensionality of the
attractors is usually very large so that their fractality is likely to
be irrelevant. This is part of the hypothesis that the system can be
regarded as an Anosov system {\it for the purpose of studying averages
of relevant quantities}. And in fact the Anosov property is used in the
above references only to obtain a representation of the SRB
distribution, \ie of the distribution describing the averages of
observables.

The same representation holds for the SRB distribution on an Axiom A
attracting set.  For this reason the fractality of an attractor was regarded
in [GC2] as "an unfortunate accident".

Therefore the {\it really non trivial hypothesis} in the mentioned
applications is the {\it reversibility of the motion on the attracting
set}.  Such reversibility is of course implied by the reversibility of
the motion on the whole phase space if the attracting set and the whole
phase space coincide: in the above references this was taken as a
consequence of the chaoticity hypothesis.

However one may wish to see how far this is justified in the cases in
which the attractor $C$ is really smaller than the whole phase space.  We
shall refer to such cases as the cases in which the attracting set
verifies Axiom A: we therefore include under the latter denomination
also the case in which the attracting set is a smooth surface (and
could therefore be said to be an Anosov system).  The possible
fractality of the closure of the attractor or its smoothness play no
role in the following.

Suppose that the {\it reversible} mechanical system under consideration
verifies Axiom C (a stronger notion than Axiom B, and a kind of "global
hyperbolicity" condition as discussed in \S2).  Suppose that the
attracting pole $\O_+={\rm clos}(C)$ {\it is not} the whole phase space
$\CC$.  Can one then conclude that the fluctuation theorem of [GC1]
holds? or at least some modification of it?

We "answer" this in the affirmative by noting that the global
time reversal map $i$, \ap assumed to exist, induces on the pole
$\O_+={\rm clos}(C)$ of the system a natural "smooth" (\ie H\"older
continuous) map $i^*$ which verifies:

$$i^* S=S^{-1}i^*,\qquad (i^*)^2=1\Eq(3.1)$$

In fact since the map $\tilde\imath$ commutes with $S$ and maps the
attracting pole $\O_+$ onto the repelling pole $\O_-$ we can set
$i^*=\tilde\imath\,i$ and define a map of $\O_\pm$ into themselves
verifying:
$$i^*S=\tilde\imath\, i \,S=\tilde\imath\,S^{-1}\, i=S^{-1}
\,i^*\Eq(3.2)$$
Hence $i^*$ is a time reversal on the future attracting set. {\it Note that
$i^*$ is not the restriction of $i$ to the future attracting set}.
This map will be the {\it local time reversal}.  One should stress that
since $\tilde\imath$ is defined only on $\O_\pm$ also $i^*$ has {\it
only} a meaning {\it as a map of such sets onto themselves}.

Its existence immediately implies the validity of a fluctuation theorem
([GC1], [GC2], [G3]) for systems that are globally reversible and
chaotic in the sense that they verify Axiom C (and hence verify the
$0$--th law and have an Axiom A attracting set), with the only
difference that the theorem applies to the phase space contraction that
occur on $\O_+$ rather than to the phase space contraction occurring in
the whole phase space $\CC$ (see \S6, (b), for a discussion).  Moreover
it implies also that the chaotic hypothesis can be conceptually
simplified.  Curiously enough this does not even require a modification
of its formulation, but it allows for a broader intrepretation of it as
the word "Anosov" can be essentially replaced by "reversible Axiom A
attracting set" and this covers explicitly the cases in which the
attracting set is not dense on phase space.

It is important to note here that the existence of $i^*$ is not trivial:
because $i^*$ {\it cannot be} (unless $\CC=\O_+$, \ie unless the
future pole of the system is the whole phase space so that the system is
actually an Anosov system) the restriction to $\O_+$ of the time
reversal map $i$, as one would naively surmise.

In fact reversible systems with Axiom A attractors usually have
attractors $C_+$ and $C_-$ for the forward motion and for the backward
motions which are {\it distinct}, [S1], in the sense that $\O_+\cap\O_-\=
{\rm clos}(C_+)\cap{\rm clos}(C_-)=\emptyset$. In such cases the time
reversal $i$ maps $\O_+$ into $\O_-$ and viceversa. Therefore although the
global time reversal $i$ has the property $Si=iS^{-1}$ it does not leave
invariant the attracting basic set $\O_+={\rm clos}(C_+)$.

The map $i^*$ is an {\it effective time reversal} acting on the closure
of the attractor $\O_+={\rm clos}(C_+)$ for the future motion.  Its
existence could be expected on philosophical grounds: if a system is
reversible there should be no way of knowing whether one is moving on
the {\it future pole} $\O_+$ or on the {\it past} pole $\O_-$.  In
particular we should be able to see that the motion is reversible
without ever even knowing about the existence of the past attractor
$C_-$.  Hence we expect that there is a "{\it local time reversal}"
$i^*$ on both the future and the past attractors: and the problem is to
find a way to construct, at least in principle, $i^*$.

The reader will notice the analogy between the above picture and the
spontaneous symmetry breaking: when the attractor dimension decreases
(because some Lyapunov exponent changes sign as a parameter changes)
the time reversal symmetry is no longer valid, but some other symmetry
survives which still has the effect of changing the sign to time: a well
known example is the breaking of $T$-symmetry in relativistic quantum
mechanics, with the $TCP$-symmetry remaining valid.

In \S4 we present a model in which $i^*$ can be constructed and
provides the paradigm of a reversible Axiom C system.  In \S5 we
discuss the meaning in symbolic dynamics of the map $i^*$.  On the
mathematical side there are various points that would require closer
investigation.  But the discussion seems to indicate that the scenario
for the construction of $i^*$ should work rather generally, as we think
it is quite naturally suggested by the results of the experiment in
[BGG].  \*

\1
\0{\it\S4. An example.} \numsec=4\numfor=1\*

We give here an example in which $i^*$, the local time reversal,
arising in the applications can be easily constructed. The example
illustrates what we think is a typical situation. The poles $\O_\pm$,
closures of an attractor $C_+$ and respectively of a repeller $\O_-$,
will be two compact regular surfaces, identical in the sense that they
will be mapped into each other by the time reversal $i$ defined below.

If $x$ is a point in $M_*=\O_+= {\rm clos}(C_+)$ the generic point of
the phase space will be determined by a pair $(x,z)$ where $x\in M_*$
and $z$ is a set of transversal coordinates that tell us how far we are
from the attractor. The coordinate $z$ takes two well defined values
on $\O_+$ and $\O_-$ that we can denote $z_+$ and $z_-$ respectively.

The coordinate $x$ identifies a point on the compact manifold $M_*$ on
which a reversible transitive Anosov map $S_*$ acts (see [G3]).  And
the map $S$ on phase space is defined by:

$$S(x,z)=(S_*x,\tilde S z)\Eq(4.1)$$
where $\tilde S$ is a map acting on the $z$ coordinate (marking a point
on a compact manifold) which is an evolution leading from an unstable
fixed point $z_-$ to a stable fixed point $z_+$.  For instance $z$
could consist of a pair of coordinates $v,w$ with $v^2+w^2=1$ (\ie $z$
is a point on a circle) and an evolution of $v,w$ could be governed by
the equation $\dot v=-\a v, \ \dot w= E-\a w$ with $\a={Ew}$.  If we
set $\tilde Sz$ to be the time $1$ evolution (under the latter
differential equations) of $z=(v,w)$ we see that such evolution sends
$v\to0$ and $w\to\pm 1$ as $t\to\pm\io$ and the latter are non marginal
fixed points for $\tilde S$.

Thus if we set $S(x,z)=(S_*x,\tilde S z)$ we see that our system is
hyperbolic on the basic sets $\O_\pm=M_*\times \{z_\pm\}$ and the
future pole $\O_+={\rm clos}(C_+)$ is the set of points
$(x,z_+)$ with $x\in M_*$; while the past pole $\O_-={\rm clos}(C_-)$
is the set of points $(x,z_-)$ with $x\in M_*$.

Clearly the two poles are mapped into each other by the map
$i(x,z)=(i^*x, -z)$.  But on each attractor a "local time reversal"
acts: namely the map $i^* (x,z_\pm)=(i^*x,z_\pm)$.

The system is "chaotic" as it has an Axiom A attracting set with closure
consisting of the points having the form $(x, z_+)$ for the motion
towards the future and a different Axiom A attracting set with closure
consisting of the points having the form $(x, z_-)$ for the motion
towards the past. In fact the dynamical systems $(\O_+,S)$ and
$(\O_-,S)$ obtained by restricting $S$ to the future or past
attracting sets are Anosov systems because $\O_\pm$ are regular
manifolds.

We may think that in the reversible cases the situation is always the
above: namely there is an "irrelevant" set of coordinates $z$ that
describes the departure from the future and past attractors. The future
and past attractors are copies (via the global time reversal $i$) of
each other and on each of them is defined a map $i^*$ which inverts the
time arrow, {\it leaving the attractor invariant}: such map will be
naturally called the {\it local time reversal}.

In the above case the map $i^*$ and the coordinates $(x,z)$ are
"obvious".  The problem is to see that they are defined quite generally
under the only assumption that the system is reversible and has a
unique future and a unique past attractos that verify the Axiom A.
This is a problem that is naturally solved in general when the system
verifies the Axiom C of \S2 (see \S3, \equ(3.1) above).

In the following section we shall describe the interpretation of $i^*$
in terms of symbolic dynamics when the system verifies Axiom C: as one
may expect the construction is simple but it is deeply related to the
properties of hyperbolic systems such as their Markov partitions.  \*

\0{\it\S5. Local time reversal and Markov partitions.}
\numsec=5\numfor=1\*

In this section we discuss the
properties of the map $i^*$ and its relation with the Markov partitions and
the symbolic dynamics.

We assume the reader familiar with the notion of Markov partition: in
any event the results of this paper logically follow those of [GC1],
[GC2] and [G2], [G3] and we can expect that only readers familiar with
those papers can have any interest in the present one.

In [GC1] we mention that a transitive Anosov reversible system admits a
Markov partition $\PP=\{Q_\s\}$, which is invariant under time reversal:
$i\PP=\PP$.

This means that for every element $Q\in \PP$ one can find an element
$Q'\in \PP$ such that $i Q= Q'$.

If the dynamical system only has a transitive Axiom A attracting set we can
still construct a Markov partition $\PP=i\PP$ but it will not have a
transitive transition matrix. The transition matrix is in fact defined
by setting $T_{\s,\s'}=1$ if $S Q_\s\cap {\rm
int}(Q_{\s'})\not=\emptyset$ (here ${\rm int}(Q)$ are the interior
points of $Q$ {\it relatively} to the closure of the attractor) and
$T_{\s,\s'}=0$ otherwise. And transitivity means that there is a power
$k$ of $T$ such that $T^k_{\s,\s'}>0$ for all $\s,\s'$.

The lack of transitivity in the above sense is simply due to the fact
that the Markov partition $\PP$ really splits into two transitive Markov
partitions, one, denoted $\PP_+$, paving the closure $\O_+={\rm clos}(C_+)$
of the future attractor and one, denoted $\PP_-$, paving the closure
$\O_-={\rm clos}(C_-)$ of the past attractor $C_-$. And of course there is no
possibility of a transition from one to the other under the action of
$S$ as the two are $S$-invariant sets.

But for Axiom C systems the Markov partition can be built in a special
way that takes into account more deeply the global time reversal
symmetry of the system.  Let in fact $O$ be a fixed point of $S$ on
$\O_+$ (if no fixed point exists $O$ can be, for the purposes of the
following discussion, be replaced by a point on one of the periodic
orbits on $\O_+$; recall that by the Axiom A property the periodic
orbits are dense on $\O_\pm$).

We shall assume, for simplicity, that $\O_+$ (hence $\O_-$ are smooth
surfaces: then we consider the stable manifold of $O$.  The latter is
dense on $\O_+$ because of the assumed transitivity of the attractor
and it has a part that is not contained on the attracting set (because we
are supposing that the attractor is {\it not} dense in phase space).

If $n_0$ is the dimension of the pole $\O_+$ and $n_s$ is the dimension
of the part of the stable manifold $W^s_O$ lying on $\O_+$ then the
dimension of the stable manifold will be ${n_s}+n$ for some $n\ge1$.
The manifold $W^s_O$ will intersect the pole $\O_-$: otherwise it would
lead to another repelling basic set, violating the assumption that
there are only two poles (\ie only one attractor for the future motion
and one for the backward motion as expressed by the $0$-th law above,
see 3) in the definition of Axiom C).

The pole $\O_-$ has (by the time reversal symmetry) the same dimension
$n_0$ of the pole $O_+$ and its intersection with $\lis W^s_O$ will be a
$n_s$-dimensional manifold in $\O_-$, an unstable manifold for the map
$S^{-1}$, dense on $\O_-$.  Likewise we can consider the point
$iO\in\O_-$ and perform the same construction by using the unstable
manifold $W^u_{iO}=iW^s_O$ of $iO$.  It will have a part of dimension
$n_u=n_s$ lying on $\O_-$ and its dimension will be $n_u+n$.  The
manifolds $W_O^s$ and $W^u_{iO}$ intersect densely on $\O_\pm$ and each
intersection point $x\in\O_+$ is on a $n$-dimensional manifold which
has one point $\tilde\imath x\in \O_-$ on $\O_-$.

It is clear that the densely defined map $\tilde\imath$ of $\O_+\otto
\O_-$ commutes with $S$ and it can be extended by continuity to a map
of $\O_+$ to $\O_-$.  If $\PP_+$ is a Markov partition of $\O_+$ then
$\tilde \imath\PP_+=\PP_-$ is a Markov partition of $\O_-$.  This is just
another way of looking at the construction of the map $\tilde\imath$,
hence of $i^*$, see \equ(3.2).

This also shows that we can establish a natural correspondence
$\s\otto\tilde\s$ between labels of elements of $\PP$ such that
$Q_{\tilde\s}=\tilde\imath Q_\s$.  Note that if $Q_\s\in \PP_\pm$ then
$Q_{\tilde\s}\in\PP_\mp$.

Note that the map $i^*$ has a very simple and natural symbolic dynamics
interpretation.  Given an allowed sequence $\V\s=\{\s_j\}$ we set
$\V{{\tilde\s}}=\{\tilde\s_{-j}\}$; since $T_{\s,\s'}=1$ means
$SQ_\s\cap {\rm int}(Q_{\s'})\not=\emptyset$, we deduce that it means
also $i\,SQ_\s\cap i\, {\rm int}(Q_{\s'})\not=\emptyset$ hence
$S^{-1}iQ_\s\cap i\,{\rm int}(Q_{\s'})\not=\emptyset$.  So that
$i\,{\rm int}(Q_\s)\cap S\,i\,Q_{\s'}\not=\emptyset$ hence
$i^*Q_{\s}\cap S\,i^*\,{\rm int}(Q_{\s'})\not=\emptyset$:

$$T_{\s,\s'}=1\otto Q_{\tilde \s}\cap S{\rm int}(Q_{\tilde
\s'})\not=\emptyset\Eq(5.1)$$
\ie $T_{\s,\s'}=1$ is equivalent to $T_{\tilde\s',\tilde\s}=1$.

This means that if $\V\s=\{\s_j\}$ is an allowed sequence of symbols for
a point $x$ on the pole $\O_+$ (\ie it is the {\it
history} on $\PP$ of a point $x\in\O_+$ in the sense that $S^j x\in
Q_{\s_j}$ for all $j$) then {\it also} $\V{{\tilde\s}}=\{\tilde\s_{-j}\}$
is an allowed sequence and $i^*x$ is the (unique) point on
the pole $\O_-$ that has $\{\tilde\s_{-j}\}$ as the history under $S$.

\*
\0{\it\S6. Markov partitions, coarse graining and trajectory segments.
Extended Liouville measure.}
\numsec=6\numfor=1\*

\0{\it a)} We first discuss the notion of coarse graining, making
precise some ideas that were advanced in [G1]. We show that Anosov
systems with Axiom A attracting sets admit, in spite of the chaoticity
of the motions that they describe, a rather natural decomposition of
phase space into cells so that the time evolution can be naturally
represented as a {\it cells permutation} and the SRB distribution can be
naturally interpreted as the distribution that gives {\it equal weight
to each of the cells}.

This may look surprising and contradictory with the property of
hyperbolicity and chaoticity of the system. Therefore it is a
particularly interesting (rather elementary) feature of the SRB
distribution which makes it even more analogous to the microcanonical
distribution in equilibrium.

Imagine that $\PP$ is a Markov partition for a transitive Axiom A
attracting set. We use it to set up a symbolic dynamic description of the
attractor.

Let $T$ be large and $\PP_T=\vee_{j=-\fra{T}2}^{\fra{T}2} S^j\PP$. Then
it is well known, [S1,R1] (see also [G2]), that we can represent the SRB
distribution as a limit of probability distributions obtained by
assigning to the elements $Q\in\PP_T$ a weight:

$$\L^{-1}_{e,T}(x_Q)\Eq(6.1)$$
where $\L_{e,T}(x)$ is the expansion rate (\ie the modulus of the
jacobian determinant) of the map $S^T$ as a map of the unstable manifold
of $S^{-T/2}x$ to that of $S^{T/2}x$, and $x_Q$ is a (suitable, see
[GC2], [G3]) point in $Q$.

Then we can imagine to partition {each} $Q\in\PP_T$ into boxes so that
the number of boxes, that we call {\it cells}, in $Q$ is proportional
to \equ(6.1).  In this way we find a representation for the SRB
distribution in which each cell of phase space has {\it the same
weight}.  The SRB distribution thus appears as the uniform distribution
on the attracting set (thus partitioned) and the time evolution can be
rather faithfully represented simply as a permutation of the cells, in
spite of the hyperbolicity.

This also shows that one has to be careful in saying that "it is obvious
that the SRB distribution is obtained by attributing the weigths
\equ(6.1) to points on the attractor", sometimes erroneously called the
"trajectory segment method": this is right only if the points are
identified with the cells of a Markov partition $\PP_T$, refinement of
a fixed Markov partition. This means that \equ(6.1) is correct only if a
suitable coarse graining of the phase space is made, and incorrect
otherwise.

If the points are chosen differently then the weight to give to each may
well be {\it very different}, as in the latter case in which it is equal
for all cells, no matter what the value of $\L^{-1}_{e,T}$ is in each of
them. And in some sense the latter representation is the most natural
one, and it realizes in general the Boltzmann idea that all points in
phase space are equivalent and the dynamics is just a one cycle
permutation of the cells, [G2].

One can say that if the system admits an Axiom A attracting set then it is
possible to define a coarse graining of phase space such that the
dynamics is {\it eventually} just a one cycle cell permutation ({\it
even though} the evolution may be non volume preserving): the SRB
distribution appears then as the {\it uniform distribution} on the
relevant phase space part (\ie the attracting set) and the (local) time
reversal as an invertible map of coarse cells onto themselves. In other
words the chaotic hypothesis can be regarded as a natural version of the
original viepoint of Boltzmann, [G1], on time evolution and ergodicity.
\*

\0b) Consider an Axiom C system. Then we can define a local time
reversal $i^*$ on the future pole. This means that a fluctuation theorem
holds for the statistics of the Liouville distribution $\m_0$ on $\CC$.
The formulation of the fluctuation theorem is unchanged provided one
defines the entropy production rate as the contraction of the surface
area on the attracting set. This is to be expected to be a rather difficult
quantity to evaluate in concrete cases because we cannot expect to have
a precise knowledge of the geometric structure of the attracting set.
In this respect one can remark that the system may have other properties
that nevertheless allow us to establish a relationship between the
contraction rate of the Liouville measure (directly accessible from the
measurement of the divergence of the equations of motion) and the
contraction rate of the surface measure {\it on the attracting set}: an
interesting instance of this has been found in [BGG], \S6, {\it(ii)},
where the extra property used was the {\it pairing rule} that held in
that case, (see [ECM1], [DM]).

In general on the pole $\O_+$ one can define a probability distribution
$\m^*_0$ that is the natural extension of the Liouville distribution in
the equilibrium case and for which the fluctuation theorem holds in the
same form that it has in the Anosov case.  It is the probability
distribution that is defined in the symbolic dynamics representation by
the {\it Gibbs distribution}, [LR], [R2], with {\it non translationally
invariant potential} given by the formal energy function:
$$H(\V \s)=\sum_{k=-\io}^{-1} h_-(\th^k\V\s)
+\sum_{k=0}^\io h_+(\th^k \V \s)\Eq(6.2)$$

\0where $\V\s$ is the symbolic sequence corresponding to a point $x$ on
$\O_+$ evaluated on a Markov partition $\PP$, see \S5; $\th$ is the
shift of the sequence $\V\s$; and we have set:
$$\eqalign{
h_-(\V \s)=&-\log \L^*_s(X(\V \s)),\quad h_+(\V \s)=\log \L_u(X(\V \s))
\cr}\Eq(6.3)$$
and $\L^*_s,\L_u$ are the jacobian determinants of the map $S$ {\it
resctricted} to the intersections of the stable or, respectively,
unstable manifolds with $\O_+$.

In the Anosov case \equ(6.1) defines a distribution equivalent to the
ordinary Liouville distribution, see [G3], [G2]. In the Axiom C case it
defines a distribution on $\O_+$ which is absolutely continuous with
respect to the surface area on $\O_+$ when the poles are smooth
manifolds (because in such case the system $(\O_+,S)$ is a Anosov system).
But if the pole $\O_+$ is just an Axiom A attracting set which is not a
smooth manifold then $\L^*_s$ is not properly defined as a jacobian
determinant (because the intersection
${\displaystyle\mathop{W}^\smile}{}^s_x=W^s_x\cap\O_+$ is
not a manifold).  Nevertheles it {\it can be defined} by using $i^*$
via:

$$\L_s^*(x)=\L^{-1}_u(i^*x)\Eq(6.4)$$
and this is our proposal for a natural extension of the defintion of
the Liouville measure on the attracting basic set $\O_+$.  It is a
distribution that may have several further properties that it seems
worth investigating.
\*
\0c) Finally, with reference to \S1 above, we note that in system like
the one studied in [BGG] it is possible that while a forcing parameter
grows the Lyapunov spectrum changes nature because some exponents
initially positive continuously evolve into negative ones as the
forcing increases.  Everytime one "positive" exponent "becomes" negative
the dimension of the future pole diminishes (usually by $2$ units when the
pairing rule is verified, see [BGG]).  At this "bifurcation" the future
pole splits into two basic sets, one will be the new pole and the other
will be its $i^*$ image.  Of course the above analysis implies that
there will be a new local time reversal $i^{**}$ on the new future
pole.  The $i^*$ image of the future pole, however, will {\it not} be
the past pole: one can easily see that the latter is more stable than
the $i$--image of the future pole: hence the past pole will be the full
time reversal of the future pole, no matter how many intermediate
bifurcations took place.

The picture in terms of diagrams, see [Sm] p.  754, is quite suggestive
and is that after $n=0,1,\ldots$ "positive" Lyapunov exponents have
"become" negative the diagram of the system consists of $2^n$ points
totally ordered starting from the past pole and going straight down to
the future pole.  During the evolution of the bifurcations $n+1$ time
reversals are defined $i^*_0\=i,i^*_1,\ldots,i^*_{n}$ and the $k$--th
time reversal $i^*_k$ leaves invariant the set of nodes in the diagram
with labels $1,2,\ldots,2^{n-k}$, $k=0,\ldots$.  \*

\0{\it Acknowledgements:} We are indebted to  E.G.D. Cohen, P. Garrido,
G . Gentile, J.L. Le\-bo\-witz for stimulating discussions and comments.
This work is part of the research program of the European Network on:
"Stability and Universality in Classical Mechanics", \# ERBCHRXCT940460,
and it has been partially supported also by CNR-GNFM and Rutgers University.

\*

\1
\0{\bf References.}
\*

\item{[CG1]} Gallavotti, G., Cohen, E.G.D.: {\it Dynamical ensembles in
nonequilibrium statistical mechanics}, Physical Review Letters,
{\bf74}, 2694--2697, 1995.

\item{[CG2]} Gallavotti, G., Cohen, E.G.D.: {\it Dynamical ensembles in
stationary states}, in print in Journal of Statistical Physics, 1995.

\item{[DM]} Dettman, , Morriss, G.: preprint 1995.

\item{[ECM1]} Evans, D.J.,Cohen, E.G.D., Morriss, G.P.: {\it Viscosity of a
simple fluid from its maximal Lyapunov exponents}, Physical Review, {\bf
42A}, 5990--\-5997, 1990.

\item{[ER]} Eckmann, J.P., Ruelle, D.: {\it Ergodic theory of strange
attractors}, Reviews of Modern Physics, {\bf 57}, 617--656, 1985.

\item{[G1]} Gallavotti, G.: {\it Ergodicity, ensembles,
irreversibility in Boltzmann and beyond}, Journal of Statistical
Physics, {\bf 78}, 1571--1589, 1995.

\item{[G2]} Gallavotti, G.: {\it Topics in chaotic dynamics}, Lectures
at the Granada school, ed. Garrido--Marro, Lecture Notes in Physics,
{\bf 448}, 1995.

\item{[G3]} Gallavotti, G.: {\it Reversible Anosov diffeomorphisms and
large deviations.}, Ma\-the\-ma\-ti\-cal Physics Electronic Journal,
{\bf 1}, (1), 1995.

\item{[G4]} Gallavotti, G.:{\it Chaotic hypothesis: Onsager reciprocity
and fluctuation--dis\-si\-pa\-tion theorem}, in {\it
mp$\_$arc@math.utexas.edu}, \#95-288, 1995. In print in Journal of
Statistical Physics, 1996.

\item{[R1]} Ruelle, D.: {\it Chaotic motions and strange attractors},
Lezioni Lincee, notes by S.  Isola, Accademia Nazionale dei Lincei,
Cambridge University Press, 1989; see also: Ruelle, D.: {\it Measures
describing a turbulent flow}, Annals of the New York Academy of
Sciences, {\bf 357}, 1--9, 1980.  For more technical expositions see
Ruelle, D.: {\it Ergodic theory of differentiable dynamical systems},
Publications Math\'emathiques de l' IHES, {\bf 50}, 275--306, 1980.

\item{[R2]} Ruelle, D.: {\it A measure associated with Axiom A
attractors}, American Journal of Mathematics, {\bf98}, 619--654, 1976.

\item{[S1]} Sinai, Y.G.: {\it Gibbs measures in ergodic theory}, Russian
Mathematical Surveys, {\bf 27}, 21--69, 1972. Also: {\it Lectures in
ergodic theory}, Lecture notes in Mathematics, Prin\-ce\-ton U. Press,
Princeton, 1977.

\item{[Sm]} Smale, S.: {\it Differentiable dynamical systems}, Bullettin
of the American Mathematical Society, {\bf 73 }, 747--818, 1967.

\item{[UF]} Uhlenbeck, G.E., Ford, G.W.: {\it Lectures in Statistical
Mechanics}, American Mathematical society, Providence, R.I.,
pp. 5,16,30, 1963.

\ciao